# QCD Confinement and the Hall D Project at Jefferson Lab

A. R. Dzierba, Indiana University, Bloomington, IN 47405, USA

Abstract

An understanding of the confinement mechanism in QCD requires a detailed mapping of the spectrum of hybrid mesons. Understanding confinement means understanding the role of gluons and it is in hybrid mesons that the gluonic degrees of freedom are manifest. High statistics searches for such states with $\pi$ and p beams have resulted in some tantalizing signals. There is good reason to expect beams of photons to yield hybrid mesons with $J^{PC}$ quantum numbers not possible within the conventional picture of mesons as $q\bar{q}$ bound states. Meager data currently exist on the photoproduction of light quark mesons. At Jefferson Lab in Newport News, VA plans are underway to upgrade the energy of the electron accelerator to 12 GeV. Along with this energy upgrade, a hermetic detector housed in new experimental hall (Hall D) will be used to collect data on photoproduced mesons with unprecedented statistics [1,2]. With 12 GeV electrons, a 9 GeV linearly polarized photon beam will be produced using the coherent bremsstrahlung technique.

## 1 QCD AND EXOTIC HYBRID MESONS

### 1.1 Overview

In the early 1970's Nambu postulated that quarks inside mesons are tied together by 'strings' in order to explain the increase of meson mass with internal angular momentum. We know now that quantum chromodynamics (QCD) describes the strong interaction between quarks. Modern lattice gauge theory (LGT) calculations show that indeed a string-like chromoelectric flux tube forms between distant static charges, as shown in Fig 1. So Nambu's conjecture was essentially correct. This flux tube leads to quark confinement and to a potential energy between the quarks that increases linearly with the distance between them. Infinite energy is needed to separate the quarks to infinity.

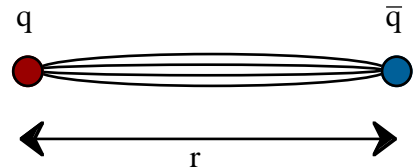

Figure 1: A flux tube forms between two static quarks.

The ideal experimental test of this feature of QCD would be to directly study the flux tube in a meson by anchoring a quark and antiquark several fermis apart and examining the flux tube that forms between them. By plucking the string one would observe two degenerate first excited states. These are the two longest wavelength vibrational modes of this system; $\pi/r$ is their excitation energy since both the mass and the tension of this "relativistic string" arise from the energy stored in its color force fields. The spectrum is shown in Fig. 2. Such a direct examination of the flux tube is of course not possible. In real life we have to be content with systems in which the quarks move. Fortunately, we know both from general principles and from lattice QCD that an approximation to the dynamics of the full system which ignores the impact of these two forms of motion on each other works quite well - at least down to the charm quark mass.

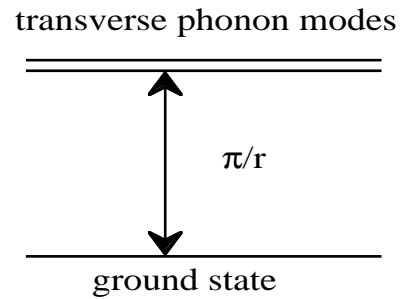

Figure 2: The spectrum of the flux tube of Fig. 1.

To extend the flux tube picture to yet lighter quarks, models are required, but the most important properties of this system are determined by the model-independent features described above. In particular, in a region around 2 GeV/$c^2$ (corresponding to the excitation energy of $\pi/r$), a new form of hadronic matter must exist in which the gluonic degree of freedom of mesons is excited. We refer to these mesons as gluonic excitations. The smoking gun characteristic of these new states is that the vibrational quantum numbers of the string, when added to those of the quarks, can produce a total angular momentum J, a total parity P, and a total charge conjugation C not allowed for ordinary $q\bar{q}$ states. These unusual $J^{PC}$ combinations, like $0^{+-}$, $1^{-+}$, and $2^{+-}$, are called exotic, and the states are referred to as exotic hybrid mesons. It is important to note that in the light quark sector hybrid mesons should form nonets just as conventional $q\bar{q}$ mesons (flux tube in the ground state) are arranged in nonets. Within the nonets we expect to find strange and non-strange mesons as well as isoscalar and isovector mesons. The unique feature of the hybrids is that exotic quantum numbers are possible. The non-exotic hybrids may mix with conventional $q\bar{q}$ making identification difficult. Establishing the hybrid nonets will depend on starting with nonets whose quantum numbers are exotic.

*1.2 Photoproduction of hybrids*

Photon beams are expected to be particularly favorable for the production of exotic hybrids. The reason is that the photon sometimes behaves as a virtual vector meson (a $q\bar{q}$ state with the quark spins parallel, adding up to total quark spin S = 1). When the flux tube in this S = 1 system is excited to its first levels both ordinary and exotic $J^{PC}$ are possible. In contrast, when the spins are antiparallel (S = 0), as in pion or kaon probes, the exotic combinations are not generated. To date, almost all meson spectroscopy in the light quark sector has been done with incident pion, kaon or proton probes or in $p\bar{p}$ annihilations. High flux photon beams of sufficient quality and energy have not been available, so there are virtually no data on the photoproduction of mesons with masses below 3 $GeV/c^2$. Thus, up to now, experimenters have not been able to search for exotic hybrids precisely where one might expect to find them.

*1.3 Evidence for gluonic excitations*

There are some tantalizing hints that gluonic excitations have been observed experimentally. Two exotic states, each with $J^{PC} = 1^{-+}$, have been reported by the E852 collaboration at Brookhaven Laboratory and confirmed in independent experiments. The first state reported has a mass of 1.4 $GeV/c^2$ and decays into $\eta\pi^-$ [3]. The interpretation of the data leading to this conclusion is not without controversy. The second state reported is perhaps on firmer ground and has a mass of 1.6 $GeV/c^2$ and decays into $\rho^0\pi^-$ [4]. In both cases the exotic signal is several percent of the more dominant signals observed in the two modes. For example in the $\eta\pi^-$ channel, the dominant signal is the $a_2(1320)$ and in the $\rho^0\pi^-$ channel the dominant signals are the $a_1(1260)$, $a_2(1320)$ and $\pi_2(1670)$. The presence of the exotic signal is not at all evident by a simple examination of the $\eta\pi^-$ or $\rho^0\pi^-$ effective mass spectrum. A partial wave analysis (PWA) must be performed. Such an analysis involves a decomposition of the mass spectrum into partial waves. The identification of a resonant wave depends on both the line shape (amplitude) and the phase motion (interference of a particular wave with other waves) as a function of mass. Application of the PWA technique to identify small signals puts stringent requirements on the detector. Exclusive events must be kinematically identified implying the need for a hermetic detector with excellent resolution and particle identification capability. With incident photons, maximum PWA information comes from using linearly polarized photons.

## 2 MAPPING OUT THE HYBRIDS

*2.1 Detector*

Fig. 3 shows a schematic layout of a hermetic detector to study the photoproduction of mesons [5]. A photon beam impinges on a 30-cm liquid hydrogen target that is located within a large superconducting solenoidal magnet. The target is surrounded by cylindrical array of scintillator strips and scintillating fibers within an outer array of straw tube drift chambers. All that is surrounded by a circular (barrel) calorimeter. Within the solenoid and downstream of the target are planar tracking chambers. Downstream of the solenoid is an atmospheric gas Cerenkov counter followed by a time-of-flight wall and then an array of lead glass that forms an electromagnetic calorimeter. The solenoidal magnet geometry is ideally suited for a high-flux photon beam. The electromagnetic charge particle background (electron-positron pairs) from interactions in the target is contained within the beam pipe by the axial field of the magnet. The solenoid already exists and will be transferred to Jefferson Lab. The lead glass array was used for E852 and is now at Jefferson Lab.

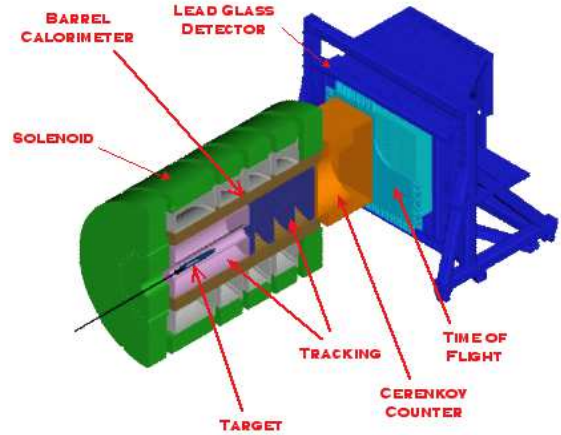

Figure 3: The proposed detector for the hybrid search at Jefferson Lab.

## 2.2 Photon beam

The photon beam will be produced using the coherent bremsstrahlung technique. At special settings for the orientation of the crystal, the atoms of the crystal can be made to recoil together from the radiating electron leading to an enhanced emission at particular photon energies and yielding linearly polarized photons. The plot of Fig. 4 shows the expected flux (in arbitrary units) of bremsstrahlung photons as a function of photon energy for an electron energy of 12 GeV. The radiator is a 20 micron-thick diamond crystal wafer. The top curve shows the energy spectrum that is a combination of incoherent radiation with its characteristic $1/E_\gamma$ fall-off (where $E_\gamma$ is the photon energy) and coherent radiation giving rise to an enhancement of flux at certain energies. The position of coherent peaks changes as one adjusts the angle of the crystal planes with respect to the electron direction. Moreover there is a correlation of the angle of the emitted photon with energy. This correlation can be exploited using collimation to reduce the incoherent background. The plot shows the resulting spectrum after collimation for a crystal orientation to produce a primary peak at 9 GeV. The average degree of linear polarization in the peak is 40 percent. In addition the photons emitted in a 0.5 GeV-wide window will be tagged using a focal plane spectrometer. The energy resolution of the photon tagger is 0.1 percent.

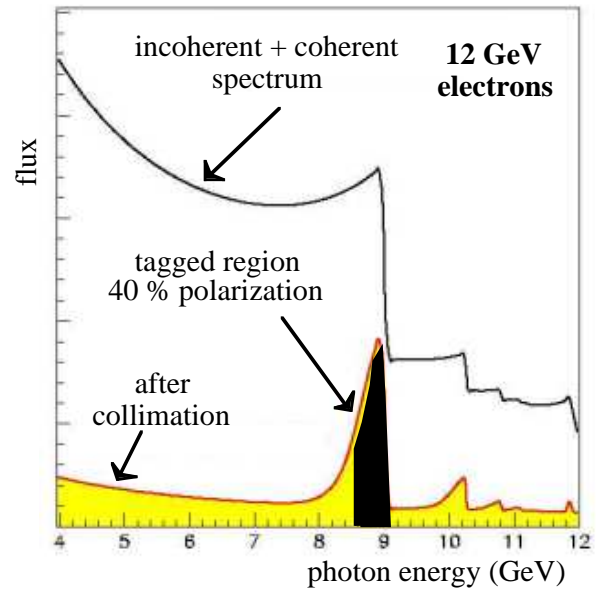

Figure 4: The flux of photons from a diamond crystal wafer with 12 GeV electrons, before and after collimation. The degree of linear polarization in the tagged region is 40 percent.

The optimal photon energy is set by several considerations. The experiment is designed to concentrate on the meson mass range from 1.0 $GeV/c^2$ up to about 2.7 $GeV/c^2$, the region where light quark hybrids are expected to exist. Incident photon energies of 9 GeV are sufficient to access this mass range. Photons of this energy will produce final state particles whose momenta and energies can be measured with sufficient accuracy with the proposed solenoidal detector. In particular, the charged particle momenta are low enough so there is no need for an additional dipole magnet that would compromise acceptance. The degree of linear polarization of photons in the coherent peak decreases as the position of the primary peak approaches the electron energy. Starting with electrons whose energy is 12 GeV, the optimal photon energy for this study is 9 GeV.

The quality of the electron beams at the CEBAF accelerator at Jefferson Lab make possible the photon beams needed to carry out the search in this proposed project. The electron beam spot size and emittance allow one to use diamond wafers and collimation. The high duty factor (essentially unity) implies high rate capability. The search for the hybrid spectrum becomes feasible with all these advances.

*2.3 Performance of the detector*

The performance of the detector and the flux and linear polarization of the photon beam determine the level of sensitivity for mapping the hybrid spectrum. A double-blind exercise was carried out in which an exotic signal, a $J^{PC} = 1^{-+}$ state with mass 1.6 GeV/$c^2$ decaying into $\rho\pi$, was generated along a mix of three well-established non-exotic states also decaying into $\rho\pi$. These non-exotic states have masses of 1.2, 1.3 and 1.7 GeV/$c^2$. In this exercise the exotic signal was generated at the level of 2.5% of the total sample – comparable to the mix observed in the E852 experiment. The momenta of the particles resulting from the decays of this mix were smeared according to the expected resolution of the detector. The experimental acceptance was also applied. The resulting dataset was passed through the PWA software. The plot of Fig. 5 shows the input exotic wave (the smooth curve) with a mass of 1.6 GeV/$c^2$ and a width of 0.170 GeV/$c^2$.

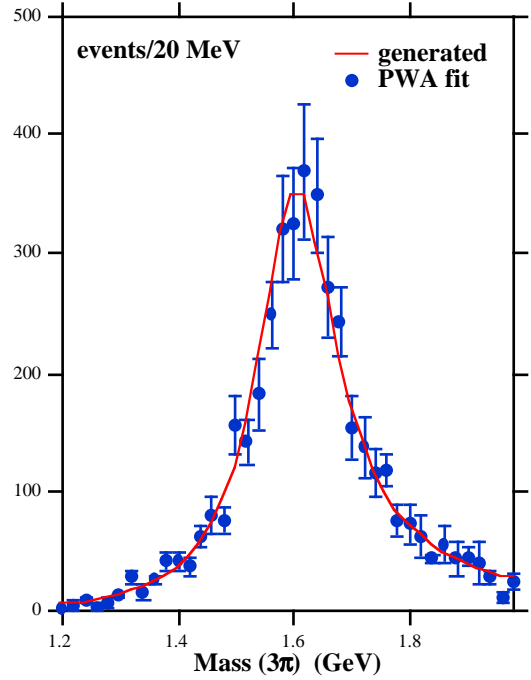

Figure 5: Results of a Monte Carlo exercise to demonstrate the performance of the detector and beam for the exotics search. The smooth curve corresponds to the input signal and the points and error bars correspond to the results after a PWA fit. Details are given in the text.

The results of the PWA fit in finite bins in mass are shown as points with error bars. The resulting fit to a Breit-Wigner resonance form yields a mass of $1.598 \pm 0.003$ GeV/$c^2$ and a width of $0.173 \pm 0.011$ GeV/$c^2$. The error bars shown in the plot correspond to statistics expected for several days of running. The establishment of the hybrid spectrum will involve a PWA for a wide variety of decay modes, of which $\rho\pi$ is one of the more straight-forward modes. PWA fits will also be done as a function of momentum transfer-squared from incident photon to outgoing meson to establish the production characteristics as well.

*2.4 Energy upgrade of CEBAF*

The CEBAF electron accelerator at Jefferson Lab employs superconducting radiofrequency (SRF) technology and its success makes possible the relatively simple and inexpensive doubling of its top energy to 12 GeV. Fig. 6 shows the layout of the CEBAF accelerator and the configuration of the proposed 12 GeV upgrade [6].

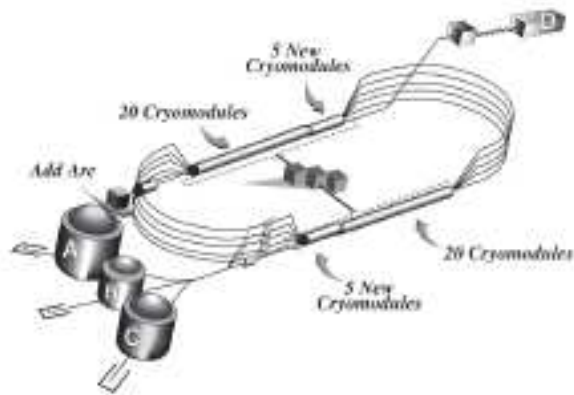

Figure 6: The configuration for the proposed 12 GeV upgrade of the CEBAF accelerator.

Each of the two linear accelerating sections currently has an empty space that can accommodate five additional SRF cryomodules. Each of these new cryomodules has 7 cells and fits in the same space as the older 5-cell modules. A trivial modification of the bending magnets in the re-circulating arcs will accommodate the higher energies. An additional arc will allow for one more pass through an accelerating section before the electrons are delivered to Hall D.

## 3 CONCLUSIONS

The design of the Hall D project is being carried out by the Hall D collaboration that currently consists of about 90 physicists from 25 institutions. The overall technique and design of the detector have successfully passed several reviews. Current plans call for a construction start for the upgrade and the Hall D project in 2004 with first physics in 2007. Data on the hybrid spectrum are essential for an understanding of confinement in QCD. The Hall D project makes use of recent developments in beam and detector technology to definitively map out the hybrid spectrum.

## 4 ACKNOWLEDGEMENTS

The author acknowledges valuable insights from Nathan Isgur. He also acknowledges his fellow Hall D collaborators including especially Curtis Meyer, Elton Smith, Adam Szczepaniak and Scott Teige.